\documentstyle[12pt]{article}
\begin{document}
\begin{titlepage}
\begin{flushright}
CINVESTAV--GRG--10\\
Mexico, July 1993
\end{flushright}
\begin{center}
\vspace{10 mm}
{\LARGE \bf Non-Abelian Cornucopions}\\
\vspace{8mm}
{{\bf E.E. Donets,}\\
{\it Dept.\ of Theoretical Physics,
Moscow State University, 119 899 Moscow, Russia,}\\
\vspace{2mm}
{\bf D.V. Gal'tsov,}\footnote{On leave from
the Dept.\ of Theoretical Physics, Moscow State
University, 119899 Moscow, Russia; e--mail: galtsov@sai.msk.su.}
\footnote{Supported by CONACyT--Mexico.} \\
{\em Dept.\ de F\'{\i}sica, Centro de Investigaci\'on y de Estudios
Avanzados del I.P.N., Apdo.\ Postal 14--740, 07000, M\'exico, D.F.,
M\'exico,}\\
\vspace{2mm}
{\bf M.S. Volkov,}\\
{\em Lab.\ of Theoretical Physics, Kazan Physical--Technical Inst.,
420 029 Kazan, Russia.}}
\end{center}
\vspace{5mm}
\begin{center}
{\bf Abstract}
\end{center}
An infinite family of cornucopions is found within the $SU(2)\times U(1)$
sector of the 4--d heterotic string low-energy theory, the extremal $U(1)$
magnetic dilatonic black hole being the lowest energy state. Non-abelian
cornucopions are interpreted as sphalerons associated with potential barriers
separating topologically distinct Yang-Mills vacua on the $U(1)$ cornucopion
background. A mass formula for non-abelian dilatonic black holes is derived,
and the free energy is calculated through the Euclidean action.\\[10mm]
PASC numbers: 04.20.Jb, 11.15 Kc, 97.60 Lf.
\end{titlepage}
\newpage

A discovery of an extremal magnetic black hole in the heterotic string theory
[1] opened new prospects in understanding the black hole evaporation puzzle
[2]. Due to miraculous cancellation of singularities in this solution the
string frame metric turns out to be completely non-singular and possessing
throat topology. An infinite volume of the throat and an associated infinite
degeneracy of the ground state gives rise to the ``cornucopion'' hypothesis [3]
as possible solution of the information problem.

A natural question arises whether the  extremal $U(1)$ magnetic dilaton black
hole (or its axionic dyon generalization [4]) is unique as cornucopion. Here
we show that, when non-abelian group is invoked,
there exists an infinite family of regular throat solutions in the string
frame.
Proliferation of cornucopions indicates that throat structure is rather
generic for the heterotic string.

Our model is the 4-dimensional coupled $SU(2)\times U(1)$
Einstein--Yang--Mills--Dilaton (EYMD) system which may be regarded as the
(truncated) bosonic part of the compactified heterotic string low-energy
theory (axion is not included because we deal with purely magnetic
configurations):
\begin{eqnarray}
I = \frac{1}{16\pi} \int d^{4}x \sqrt{-g} \{l_{Pl}^{-2}[-R +
2(\nabla \phi)^2] -  \nonumber\\
- e^{-2\phi}[2Tr F^2 + {\cal F}^2]\} +
\frac{1}{8\pi l_{Pl}^2}\oint (K-K_{0})d^{3}\Sigma .
\end{eqnarray}
Here $\phi$ is the dilaton, $F=dA-ig[A,A]$ is the $SU(2)$ gauge field,
$A=\frac{1}{2}\tau^{a}A^{a}_{\mu}dx^{\mu}$ ($\tau^{a}$ are  Pauli
matrices and $g$ is the $SU(2)$ gauge coupling constant); ${\cal F}=d {\cal
A}$ is the $U(1)$ field.

We use curvature coordinates to parameterize the spherically symmetric static
space-time:
\begin{equation}
ds^{2} = \frac{l^{2}_{Pl}}{g^{2}}
[(1-\frac{2m}{r})\sigma^{2}dt^{2} -
\frac{dr^{2}}{1-2m/r} - r^{2}d \Omega] ,
\end{equation}
where and $m$ and $\sigma$ are two
functions of radial variable. Asymptotic flatness implies
$m \rightarrow M $, (ADM mass), $\sigma \rightarrow 1$, and
$\phi = \phi_{0} + D/r + O(r^2)$, as $r \rightarrow
\infty$, where $D$ is the dilaton charge.

The magnetic $SU(2)$ connection in the gauge $A_{0}=0$ reads [5]
\begin{equation}
gA = (f-1)(L_{3}d\vartheta - \nu \sin\vartheta L_{2}d\varphi),
\end{equation}
where $L_{2}=\partial_{\vartheta}L_{1}$, $L_{3}=(\nu \sin\vartheta)^{-1}%
\partial_{\varphi}L_{1}$, $2L_{1} =\tau_1 \sin\vartheta \cos\nu\varphi  +%
\tau_2 \sin\vartheta \sin\nu\varphi + \tau_3 \cos\vartheta$ , $\nu$ is an
integer and $f$ is a real function of $r$. We also assume the $U(1)$ magnetic
monopole field $g{\cal F} = q \sin\vartheta d\vartheta \wedge
d\varphi$ with constant $q$.

The field equations are
\begin{equation}
(\sigma e^{-2\phi}\Delta f'/r^2)' = \sigma e^{-2\phi}
f(f^{2}-1)/r^2,
\end{equation}
\begin{equation}
(\Delta\sigma\phi')' + \sigma F e^{-2\phi}=0,
\end{equation}
\begin{equation}
2m' = \Delta \phi'^2 + F e^{-2\phi},
\end{equation}
\begin{equation}
(\ln \sigma)' =G \equiv  r\phi'^{2} + 2e^{-2\phi}f'^{2}/r,
\end{equation}
where $\Delta= r^2 - 2mr$ and $r^2F= 2\Delta f'^2 + \nu^2(f^2-1)^2 + q^2$.
Eq.\ (4) is applicable only in the case $\nu^{2}=1$. If
$\nu^{2}\neq 1$, an additional YM constraint appears:
$f'=0$, which together with (4) (acquiring an extra numerical factor)
yields $|f(r)|\equiv 0, 1$.
For $\nu^{2}=1$, more general solutions with $f\neq const$ exist too,
in this case $ |f(\infty)|= 0, 1$ to assure asymptotic
flatness [9]. For $q=0$ Eqs.\ (4--7) reduce
to those of the $SU(2)$ EYMD theory [5--7].

This system admits a remarquably simple first integral
\begin{equation}
\log[(1-\frac{2m}{r})\sigma^{2}] + 2(\phi - \phi_{0})=
C\int_{r}^{\infty}\frac{dr}{\sigma r(r-2m)},
\end{equation}
where $C$ is an integration constant. The integrand in the right hand side
has poles at $r=0, r=2m(r)$ and at (possible) zeros of $\sigma
(r)$, so the integral diverges if the integration region contains any of
the poles. If the left hand side of (8) is regular on the semiaxis
(though not necessarily each of two terms), then $C=0$ and therefore
\begin{equation}
g_{00} =\exp(-2(\phi -\phi_{0})).
\end{equation}
This is exactly the condition assuring a``synchronous'' form of the
string metric
\begin{equation}
d\tilde{s}^{2} = \exp(2\phi)ds^{2} = d\tau^{2} -h_{ij}(x^{k})dx^{i}dx^{j},
\end{equation}
with $d\tau = \exp(\phi_{0})dt$. It also implies
the equality of the dilaton charge to the ADM mass $ D=M$.

One may wonder when the identity (9) can hold.
Consider first an abelian ($f\equiv 0, \pm1$) 4-parametric solution
\begin{equation}
\sigma=(1+\frac{D^2}{r^2})^{-1/2},\,
\Delta =\frac{\sigma(r^2+p^2)-2Mr}{\sigma^3},\,
e^{-2(\phi-\phi_0)}=\frac{1-\sigma}{1+\sigma},
\end{equation}
where $p=P\exp(-\phi_{0})$, $P$ being the magnetic charge,
$M=\frac{1}{2}\sqrt{r^{2}_{H}+2p^{2}}$ is the Schwarzschild mass ,
$r_{H}\geq 0$ is the radius of an event horizon, $g_{00}(r_{H})=0$, and
$D=p^{2}/2M$ is the dilaton charge. For solutions with $f\equiv 0$
the charge $P=\sqrt{\nu^{2}+q^{2}}$ contains contributions
of both $U(1)$ and $SU(2)$ sectors; if $|f|\equiv 1$,
$P=q$ is purely $U(1)$. The metric (11) can be put into the Garfinkle
et al.\ [1] form by a suitable coordinate transformation.

Substituting (11) into (8) one can see that $C=0$ if and only if
$r_{H}=0$, i.e. in the extremal limit. For $r_{H}\neq 0$ the logarithm is
divergent, while the $\phi$-term is finite, so $C\neq 0$ to
compensate the divergence. Clearly, any other solution with
$\sigma (r_{H})\neq 0, \phi (r_{H}) < \infty$ will demand $C\neq 0$. An
important example is provided by non-abelian dilatonic black holes
[5--7] which have regular event horizon of a finite radius.

Vanishing of $r_{H}$ may correspond
either to  {\em globally regular solutions} [5--6], or to
cornucopions. For globally regular solutions the equality
(9) is inevitable, since $m(0)=0, \sigma (0)\neq 0$,
$\phi (0)$ is finite, while the integral at the right hand side of the
(8) diverges as $r\rightarrow 0$, what implies $C=0$.
Such solutions exist for $\nu^{2}=1$, form an infinite
sequence labelled by the number $n$ of zeros of the YM function $f_n$,
($|f_n(\infty)|=1$), and possess discrete values of mass $M_{n}$ and
dilaton charge $D_{n}$ increasing with $n$. Numerically $M_{n}$ and $D_{n}$
were found to be equal with an accuracy of the order of $10^{-3}$ [5].
Now it is clear from (9) that this equality actually is exact.

Consider now the case of a shrinking horizon. We want to show that there exists
an infinite family of hybrid structures which exhibit features both of the
extremal $U(1)$
magnetic dilaton black holes and regular $SU(2)$ solutions. Let us
seek a solution which generates both terms in the left hand side of (8)
singular as $r\rightarrow 0$, but $C=0$, i.e. divergencies mutually
cancel. Locally
\begin{eqnarray}
m=r(3-kr^{2})/8 + O(r^{5}),\nonumber \\
\sigma e^{-\phi_0} = \sqrt{2}r(1-kr^{2})/q + O(r^{5}), \nonumber \\
f = 1+ br^{2} - (k+3b^{2})br^{4}/4 + O(r^{6}),\nonumber \\
e^{-2\phi} = r^{2}(1-kr^{2})/2q^2 + O(r^{6}),
\end{eqnarray}
where $b$ and $k$ are free parameters and we set $\nu=1$. The abelian solution
(11) with $|f|\equiv 1$ corresponds to $b=0$ and $k=q^{-2}\exp(2\phi_0)$.
However, numerical integration of (4--7) shows that the expansions
(12) can be matched with desired regular asymptotics for many other values of
$b$ and $k$. The resulting solutions are qualitatively similar to a particular
set found recently in the
$SU(3)$ EYMD theory [8]\footnote{From the Eq.\ (9) of the present
paper it is clear that in [8] $\sigma_{0}=2\sqrt{2}/\sqrt{3}$, so that the
string metric (28) is exactly synchronous.}. The parameters $b$ and $k$
form discrete sequences $b_{n}(q), k_{n}(q)$, increasing  with $n$.
The function $f$ oscillates $n$ times around zero and goes
asymptotically to $(-1)^{n+1}$. Dilaton field monotonically
decreases, while the mass function $m(r)$ increases up to the asymptotic
value $M_{n}$. The sequence $M_{n}=D_n$ grows up with $n$, the lowest
$n=0$ member being $M_{0}=q\exp(-\phi_{0})/{\sqrt{2}}$ --- the mass
of the $U(1)$ cornucopion. Numerical values of $b$ and $k$ for $\phi_0=0$
and $n=0, 1, 2, 3$ are given below ($n=0$ corresponds to the
abelian solution (11)).
\begin{center}
\begin{tabular}{||l||l|l|l||l|l|r||} \hline
& \multicolumn{3}{c||}{$q^2=1$} & \multicolumn{3}{c||}{$q^2=10$} \\
\cline{2-4} \cline{5-7}
$n$ &  $k$  &  $b$  &  $M$  &  $k$   &   $b$   &   $M$    \\ \hline
0   & 1     &  0    & 0.707 &  0.1   &   0     &  2.236   \\
1   & 2.37  & 0.592 & 0.942 &  0.121 & 0.093   &  2.323   \\
2   & 10.6  & 4.674 & 0.990 &  0.250 & 0.736   &  2.341   \\
3   & 61.1  & 29.95 & 0.998 &  1.046 & 4.713   &  2.344\\ \hline
\end{tabular}
\end{center}

For all $n$ the string frame interval in the throat is
$d\tilde{s}^{2} = d\tau^{2} - 8q^2(d\ln r)^2 - 2q^2 d\Omega$
with the same $\tau$ as in (10). Behaviour of $f$ outside the throat is
qualitatively similar to that of the EYM black holes [9,10]. In particular, the
extremal EYM black holes [10] have $|f|=1$ at the horizon too. An important
difference, however, is that in the charged EYM case the extremality limit
generally corresponds to some {\em finite} $r_{H}$. In presence of a dilaton
the metric function $\Delta$ can not have double zero unless $r_H=0$, and
we have solutions which outside the
throat are similar rather to regular EYM sphalerons [11], than to black holes.

In view of this remark it is tempting to interpret non-abelian cornucopions
as sphalerons separating topologically distinct YM vacua on the $U(1)$
cornucopion background. Our argument here is similar to given in [11]. Define
a set of YM  $k$-vacua in the U(1) cornucopion sector as
$\{g_{\mu \nu}, {\cal A}_{\mu}, \phi,
A_{\mu}^{(k)}=i g^{-1} U_k\partial_{\mu} U_k^{-1}\}$,
where $g_{\mu\nu},{\cal A}_{\mu},\phi$ correspond to the $U(1)$ cornucopion
and $A_{\mu}^{(k)}$ is the pure gauge YM connection with winding number $k$.
Then one can construct a path in the function space of the
$SU(2)\times U(1)$ EYMD theory connecting two YM vacua:
$\{g_{\mu\nu}(\lambda ), {\cal A}_{\mu}(\lambda ), \phi(\lambda),%
A_{\mu}(\lambda )\}$, where $\lambda\in [0,\pi]$, and
\begin{equation}
A_{\mu}(\lambda )=i\frac{1-K(r)}{2g}U\partial_{\mu}U^{-1}, \; %
U=\exp(2i\lambda L_{1}),
\end{equation}
imposing boundary conditions $ K(0)=1, K(\infty )=-1$. The abelian component
is assumed to be constant along the path
${\cal A}_{\nu}(\lambda )={\cal A}_{\nu}(\lambda =0)$,
the dilaton is defined through (9) all over
the path, and the metric $g_{\mu\nu}(\lambda)$ is assumed to be of the form
(2) with the functions $m(\lambda )$ and $\sigma(\lambda )$
resulting from the (off shell) $G^{0}_{0}$ and $G^{r}_{r}$ Einstein equations
with $\lambda$-dependent stress-energy tensor.
The resulting path is the family of configurations possessing the same
behavior near the origin and the same total charges as the initial
vacuum state. It meets YM vacua at both ends
($\lambda=0, \pi$). If we put in (13) $K(r)=f_n(r)$ where $f_n$ is
the $n$-th non-abelian cornucopion solution (with odd $n$), the point
$\lambda=\pi/2$ will give exactly the $n$-th non-abelian solution.
Taking $\lambda$ to be time-dependent, so that $\lambda (t=-\infty)=0$ and
$\lambda (t=\infty )=\pi$, and gauging out an appearing $A_{0}$,
one can realize that this path connects a
trivial vacuum with zero winding number and a neighboring one with
unit winding number [11]. We conclude that non-abelian cornucopions
lie at stationary points of potential barriers separating topologically
distinct YM vacua in the $U(1)$ cornucopion sector. (The structure of the
energy surface around stringy sphalerons is rather different from that of
the electroweak sphaleron, details will be given elsewhere.)

In conclusion we discuss thermodynamics of non-abelian black holes in
presence of a dilaton. The imaginary time prescription
gives the following general expression for the Hawking temperature
\begin{equation}
T=\sigma (r_{H})\Delta'(r_H)/4\pi r^2_{H},
\end{equation}
which is valid for any black hole solution such that $\Delta$ has
simple zero at $r=r_H$. If $\Delta$ has double zero (extremal solution),
this expression gives vanishing $T$, unless $r_H=0$. However, in our case just
$r_H \rightarrow 0$ and $\sigma (r_H) \rightarrow 0$ too. Amazingly,
(14) still holds in this limit
and gives the correct (non-zero) $T$, what can be checked by applying
the imaginary time argument directly to the expansion (12).

To derive the first law of thermodynamics, we express the mass $M$
through Einstein equations (6--7) as a functional of $f$ and $\phi$
depending on three parameters $r_H$, $q$ and (implicitly) $\phi_0$ :
\begin{equation}
M=\frac{r_{H}}{2}\exp(-\int_{r_{H}}^{\infty}G dr) +
\frac{1}{2} \int_{r_H}^{\infty}dr(r^2\phi'^2 + Fe^{-2\phi})
\exp(-\int_r^{\infty}Gdr).
\end{equation}
Its on-shell variation over the parameters $r_H$, $q$ and $\phi_0$ yields
\begin{equation}
\delta M=T\delta S+\mu_{q}\delta q - D\delta\phi_{0},
\end{equation}
where $S=\pi r_{H}^{2}$ has the meaning of the entropy, $T$ is the
temperature (14), and
\begin{equation}
\mu_{q} = q\int_{r_{H}}^{\infty}\frac{\sigma}{r^{2}}e^{-2\phi}dr
\end{equation}
is the chemical potential for the magnetic charge. Hence the entropy
of the (either abelian or non-abelian) dilatonic black hole is,
as usual, the quarter of the event horizon area in Planck's units.
Note, that the dilaton charge and the asymptotic value $\phi_{0}$ enter as
thermodynamically conjugate quantities. Integrating (5)
we obtain for the dilaton charge
\begin{equation}
D=\int_{r_H}^\infty\sigma F e^{-2\phi}dr.
\end{equation}
\noindent
In the abelian case $|f|\equiv1$ one has $D=q\mu_q$, and two
last terms in (16) combine to yield
$\delta M= T \delta S +\bar{\mu} \delta \bar{q}$,
where $\bar{q}=q\exp(-\phi_0)$, in agreement with Ref.\ [12].

To check the consistency of these considerations let us calculate the
Euclidean action $I_{E}=I_{G}+I_{m}$. Putting
$\tau =it$ and using (1) and (2) we get for the gravitational part
\begin{equation}
I_{G}=\frac{1}{T}\int_{r_{H}}^{\infty}m\sigma 'dr +
\left. \frac{1}{2T}[\sigma(mr)'-\sigma '\Delta]\right| _{r=r_{H}},
\end{equation}
where the horizon term comes from the total derivative in $\int R\sqrt{-g}$ .
The matter part simplifies on shell as
\begin{equation}
I_{m}=\frac{1}{2T}\int_{r_{H}}^{\infty}dr\sigma%
(\Delta \phi'^2 + Fe^{-2\phi})=
\frac{1}{T}\int_{r_{H}}^{\infty}dr\sigma m'.
\end{equation}
The first term in (19) combines with (20) to give a total
derivative and finally we get
$I_{E}=F/T$, where $ F=M-TS$ is the free energy.
Thus, the field-theoretical definition of the entropy
gives the same geometric value $S=\pi r_H^2$. This is in harmony with the
result of Kallosh et al.\ [13] obtained within an abelian model.
In the cornucopion limit $r_{H}\rightarrow 0$,
with fixed $q\neq 0$, the entropy goes to zero, while the temperature remains
finite and equal to the abelian value
\begin{equation}
T_n=T_0=\frac{e^{\phi_0}}{4\sqrt{2} \pi q}=\frac{1}{8\pi M}.
\end{equation}
In contrary, the chemical potential (17), being dependent globally on
$\sigma$ and $\phi$, takes different values for different $n$.

To summarize, we have shown that the metric-dilaton identity (9) which assures
regular throat structure of the stringy space-time admits infinitely many
non-abelian realizations. The existence of new non-abelian cornucopions is
related to YM topology and indicates that potential barriers separating
different topological sectors have finite heights. So far both regular [5]
and cornucopion sphalerons have been obtained using the low-energy effective
action. They are not self-dual and are likely to be affected by
higher order corrections. Nevertheless, from the above topological argument
it seems plausible that (suitably renormalized) sphaleron solutions will
persist

in the full heterotic string theory too.

One of the authors (D.V.G.) highly appreciates the hospitality of the
Department of Physics of CINVESTAV, Mexico, and the financial support by
CONACyT. The work was also supported in part by the Russian Foundation for
Fundamental Research (project Fi--93--02--16977).

\end{document}